\documentclass[prd,twocolumn,amsmath,amssymb,floatfix,superscriptaddress]{revtex4-1}

\usepackage{graphicx}
\usepackage{amssymb}
\usepackage{amsmath}
\usepackage{hyperref}

\usepackage{color}

\DeclareMathAlphabet\mathbfcal{OMS}{cmsy}{b}{n}
\definecolor{darkgreen}{cmyk}{0.85,0.2,1.00,0.35} 
\definecolor{purple}{cmyk}{0.5,1.0,0,0}

\newcommand{\stucky}{St\"{u}ckelberg}
\newcommand{\parg}{}
\newcommand{\Gs}{\delta \Gamma}
\newcommand{\mom}{\delta q}
\newcommand{\tr}[1]{[#1]}
\newcommand{\fa}{a_F}
\newcommand{\fadot}{\dot a_F}
\newcommand{\pr}{\delta T^r_{\hphantom{r}r}}
\newcommand{\pt}{\delta T^\theta_{\hphantom{r}\theta}}
\newcommand{\pp}{\delta T^\phi_{\hphantom{r}\phi}}
\newcommand{\Tt}{\tau}
\newcommand{\dfeom}{\Gs_{\rm eom}}

\newcommand{\ta}{{a}}
\newcommand{\tX}{{X}}
\newcommand{\gisig}{ {\boldsymbol{\Sigma}}}

\def\barray{\begin{array}}
\def\earray{\end{array}}
\def\be{\begin{equation}}
\def\ee{\end{equation}}
\def\ben{\begin{equation} \nonumber}
\def\een{\end{equation}}
\def\ban{\begin{eqnarray*}}
\def\ean{\end{eqnarray*}}
\def\ba{\begin{eqnarray}}
\def\ea{\end{eqnarray}}

\def\({\left(}
\def\){\right)}
\def\half{{1\over2}}

\begin{document}

\title{Self-accelerating Massive Gravity:\\ Time for Field Fluctuations}
\author{Mark Wyman}
\affiliation{Kavli Institute for Cosmological Physics, Department of Astronomy \& Astrophysics,  Enrico Fermi Institute, University of Chicago, Chicago, Illinois 60637, U.S.A}
\author{Wayne Hu}
\affiliation{Kavli Institute for Cosmological Physics, Department of Astronomy \& Astrophysics,  Enrico Fermi Institute, University of Chicago, Chicago, Illinois 60637, U.S.A}
\author{Pierre Gratia}
\affiliation{Department of Physics, University of Chicago, Chicago, Illinois 60637, U.S.A}
\begin{abstract}
The ghost-free theory of massive gravity has exact solutions where the effective stress energy generated
by the graviton mass term is a cosmological constant for any isotropic metric. Since they are exact, these solutions mimic a cosmological constant
in the presence of any matter-induced isotropic metric perturbation. In the \stucky\ formulation, this stress energy
is carried entirely by the spatial \stucky\ field.  We show that any stress energy carried by fluctuations in the spatial field away from the exact solution
always decays away in an expanding universe.  
However, the dynamics of the spatial \stucky\ field perturbation depend on the background
 temporal  \stucky\ field, which is equivalent to the unitary gauge time coordinate.  This dependence resolves an 
 apparent conflict in the existing literature by showing that there is a special unitary time choice for which the field dynamics and energy density perturbations vanish identically.  In general, the isotropic system has a single dynamical degree of freedom requiring two sets of initial data; however,
 only one of these initial data choices will affect the observable metric.  Finally, we construct cosmological solutions with
 a well-defined perturbative initial value formulation and comment on alternate solutions that evolve
 to singularities.
\end{abstract}

\maketitle
\section{Introduction}
Constructed  to remove the Boulware-Deser ghost \cite{Boulware:1972zf},
the theory of massive gravity  \cite{Gabadadze:2009ja,deRham:2009rm,deRham:2010ik,deRham:2010kj} also possesses solutions that accelerate the
 cosmological expansion in the
absence of a true cosmological constant  \cite{deRham:2010tw,Koyama:2011xz,Koyama:2011yg,Nieuwenhuizen:2011sq,Berezhiani:2011mt,D'Amico:2011jj,Gumrukcuoglu:2011ew,Gratia:2012wt,Kobayashi:2012fz,Volkov:2012cf,Volkov:2012zb}. In a previous paper  \cite{Gratia:2012wt}, we demonstrated that
all of these apparently distinct solutions are part of a single class of isotropic solutions
in the \stucky\ formulation of the theory.    This class possesses  several key features. 
The spatial \stucky\ fields follow the radial coordinate of the metric itself and  produce the same effective
cosmological constant in the presences of any isotropic metric.  
The temporal \stucky\ field is inhomogeneous in isotropic coordinates but
does not play a direct role in establishing the effective cosmological constant.   
Solutions for this field are not uniquely specified,  leading to seemingly distinct versions of self-acceleration. 

{Once the \stucky\ fields are set on a self-accelerating solution, they remain on it for
any evolution of the matter fields that remains isotropic, including cosmological expansion
and radial collapse of matter perturbations.   }
Given the inherent interest of self-accelerating cosmological solutions, it is important to understand whether \stucky\ field perturbations around them
are themselves stable and healthy. This question has now been widely studied, with different authors using different approaches.
Confusingly, these different approaches seem to give different conclusions. Study of perturbations in a decoupling
limit showed a potentially healthy scalar degree of freedom but potentially problematic (strongly coupled or ghost-like) vector
degrees of freedom \cite{deRham:2010tw,Koyama:2011wx, Tasinato:2012ze}. A local patch expansion approach, which one might
expect to be qualitatively similar to a decoupling limit, found no propagating \stucky\ degrees of freedom at all \cite{D'Amico:2012pi}. 
Finally, study of perturbations in the full theory around a particular solution for the temporal \stucky\ field similarly found
no propagating  degrees of freedom \cite{Gumrukcuoglu:2011zh}\footnote{We note for clarity that these considerations are distinct from the question
of the regime of validity of massive gravity considered as an effective theory; the theory as written is known to be strongly coupled above a scale related to
$\Lambda_3=(M_{\rm pl} m^2)^{1/3}$ (where $M_{\rm pl}$ is the Planck mass and $m$ the graviton mass)  (e.g. \cite{deRham:2010gu}).}.

{To resolve this issue, we study spherically symmetric \stucky\ perturbations around the full theory for the whole class of solutions.  We find that in general the spatial \stucky\ field fluctuation does possess dynamics
and carries stress-energy fluctuations in addition to the background constant.
There are special choices
of the temporal \stucky\  field background that eliminate the dynamics but also eliminate the possibility of energy density fluctuations.  In all cases, isotropic energy
density fluctuations are stable and from general initial conditions decay back to the cosmological constant
in an expanding universe.
Our approach allows us to show that the decoupling limit findings \cite{deRham:2010tw,Koyama:2011wx} and specific temporal solutions in the 
full theory \cite{Gumrukcuoglu:2011zh} are not, after all, in conflict. We have not been able to 
harmonize our findings with Ref.~\cite{D'Amico:2012pi}, who finds no dynamics 
for either temporal \stucky\ field choice. {Because we restrict ourselves to isotropic perturbations, we are not able to address the generality of recent calculations \cite{DeFelice:2012mx,Gumrukcuoglu:2012aa} that demonstrate that at least some of these solutions are unstable to anisotropic perturbations.}

The structure of this paper is as follows.  In \S \ref{sec:acceleration}, we   briefly review the theory of massive gravity \cite{deRham:2010kj} and the class of isotropic self-accelerating solutions \cite{Gratia:2012wt}. We  then consider the equations of motion and action governing spherically symmetric field perturbations around this
class of solutions in \S \ref{sec:fluctuations}.   Finally in \S \ref{sec:examples}  we show how the choice of the temporal \stucky\ field in the
background affects the dynamics of the perturbations and their contribution to
stress energy.   
We close with a discussion and summary of these results in \S \ref{sec:discussion}.

\section{Self-Acceleration}
\label{sec:acceleration}

In the section, we review the action, equations of motion, and stress energy
for  the massive gravity model 
\cite{deRham:2010ik,deRham:2010kj}.  We specialize these quantities for the
 class of exact isotropic self-accelerating solutions \cite{Gratia:2012wt} which will form the basis of the perturbation studies that follow.

\subsection{Massive Stress Energy}

The  Lagrangian density \cite{deRham:2010ik,deRham:2010kj}
\begin{align}
\mathcal{L}_G &=\frac{M_{\rm pl}^2}{2}\sqrt{-g}\left[ R-\frac{m^2}{4}\mathcal{U}(g_{\mu\nu},\Sigma_{\mu\nu})\right],
\end{align}
represents a  covariant theory of massive gravity
constructed so as to eliminate the Boulware-Deser ghost \cite{Boulware:1972zf, Hassan:2011hr, Hassan:2011ea}.
Here $m$ is the graviton mass and $M_{\rm pl}$ is the reduced Planck mass. The potential $\mathcal{U}$ is constructed in a covariant manner
by making it a function of the so-called fiducial metric $\Sigma_{\mu\nu}$
\begin{equation}
\Sigma_{\mu\nu} = \partial_{\mu}\phi^a\partial_{\nu}\phi^b\eta_{ab},
\end{equation}
where  $\phi^a$  are the 4  \stucky\ fields.  These fields transform as spacetime scalars and
hence maintain general covariance.

The potential $\mathcal{U}$ can now be written in terms of the matrix
\begin{equation}
\Sigma^{\mu}_{\;\,\nu} \equiv g^{\mu\alpha}\Sigma_{\alpha\nu},
\label{eqn:potentialmatrix}
\end{equation}
as
\begin{align}
{\mathcal U \over 4} &=
-12+6\tr{\sqrt{\gisig}}+\tr{\gisig}-\tr{\sqrt{\gisig}}^2 
\nonumber\\&\quad
+\alpha_3 \Big( {-24} + 18\tr{\sqrt{\gisig}} - 6\tr{\sqrt{\gisig}}^2 + \tr{\sqrt{\gisig}}^3
\nonumber\\&\qquad
- 3 \tr{\gisig} (\tr{\sqrt{\gisig}}-2) + 2 \tr{\gisig^{3/2}}\Big)
\nonumber\\&\quad
+\alpha_4 \Big( {-24}+  24 \tr{\sqrt{\gisig}} -12 \tr{\sqrt{\gisig}}^2- 12 \tr{\sqrt{\gisig}} \tr{\gisig} 
\nonumber\\&\qquad
+ 6\tr{\sqrt{\gisig}}^2 \tr{\gisig}+ 4\tr{\sqrt{\gisig}}^3 + 12  \tr{\gisig} - 3 \tr{\gisig}^2\nonumber\\&\qquad
 - 8  \tr{\gisig^{3/2}}
(\tr{\sqrt{\gisig}}-1)
+ 6\tr{\gisig^2} -  \tr{\sqrt{\gisig}}^4 \Big),
\label{eqn:pottrace}
\end{align}
where brackets denote traces, $\tr{{\bf A}} \equiv A^\mu_{\;\,\mu}$, and $\alpha_3$, $\alpha_4$ are 
free parameters.

Variation of the action with respect to the metric yields the modified Einstein
equations
\begin{equation}
G_{\mu\nu} =  m^2 T_{\mu\nu} + \frac{1}{M_{\rm pl}^2} T_{\mu\nu}^{(m)},
\label{eqn:modeinstein}
\end{equation}
where $G_{\mu\nu}$ is the usual Einstein tensor and $T_{\mu\nu}^{(m)}$ is the matter stress energy tensor.  Here 
\begin{align}
T_{\mu \nu}& =  \frac{1}{\sqrt{-g}} \frac{\delta}{\delta g^{\mu \nu}}
\sqrt{-g} \, \frac{\mathcal U } {4} 
\label{eqn:stressenergy}
\end{align}
is the dimensionless effective stress energy tensor provided by the mass term.
An explicit expression in terms of $\gisig$ is given in Ref.~\cite{Gratia:2012wt}.

\subsection{Effective Cosmological Constant}

A constant stress energy
is an exact solution of massive gravity 
for any 
spatially isotropic metric \cite{Gratia:2012wt},
\begin{equation}
ds^2=-b^2(r,t) dt^2+\ta^2(r,t)(dr^2+r^2d\Omega^2).
\label{eqn:metric}
\end{equation}
To see this fact, consider that an isotropic parameterization of the 
 \stucky\ fields
\begin{align}
\phi^0 &= f(t,r) ,\nonumber\\
\phi^i &= g(t,r) \frac{x^i}{r} ,
\end{align}
enables the potential to be written more compactly as
\begin{equation}
{\frac{\mathcal U}{4}} = P_0\left( \frac{g}{\ta r} \right) + \sqrt{\tX}P_1\left( \frac{g}{\ta r} \right)
+  W P_2\left( \frac{g}{\ta r} \right),
\label{eqn:genpot}
\end{equation}
where 
\begin{align}
\tX & \equiv\Bigl(\frac{\dot{f}}{b}+\mu\frac{g'}{\ta }\Bigr)^2-\Bigl(\frac{\dot{g}}{b}+\mu\frac{f'}{\ta }\Bigr)^2, \nonumber\\
W & \equiv \frac{\mu}{ab} \( \dot f g' - \dot g f' \),
\label{eqn:XW}
\end{align}
and $\mu\equiv{\rm sgn}(\dot f g' - \dot g f')$.  
Here and below overdots denote derivatives
with respect to $t$ and primes  denote derivatives with respect to $r$ when acting on \stucky\ or metric fields.
{Note that $W$ is related  to the determinant
of ${\bf \Sigma}^{1/2}$ which we assume is never zero.  Correspondingly $\mu$ must
be either $+1$ or $-1$ throughout the spacetime.   This will be an important consideration
in \S \ref{sec:examples}.}

The $P_n$ polynomials are
\begin{align}
P_0(x) &= - 12 - 2 x(x-6) - 12(x-1)(x-2)\alpha_3 
\nonumber\\&\qquad -24(x-1)^2\alpha_4 ,\nonumber\\
P_1(x) &= 2 (3 -2 x)  +  6(x-1)(x-3)\alpha_3 +   24(x-1)^2 \alpha_4,\nonumber\\
P_2(x) &= -2 + 12 (x-1) \alpha_3 - 24(x-1)^2 \alpha_4,
\end{align}
and satisfy the recursion 
\begin{equation}
P_n' = 2 P_{n+1} - x P_{n+1}' ,
\label{eqn:recursion}
\end{equation}
where here and throughout $P_n'(x)\equiv dP_n/dx$ and should not be confused with radial
derivatives.

Varying the action with respect to $f$ and $g$ yields the \stucky\ field equations
\begin{align}
\label{eqn:eomf}
&\partial_t\Biggl[\frac{\ta ^3r^2}{\sqrt{\tX}}\Bigl(\frac{\dot{f}}{b}+\mu\frac{g'}{\ta }\Bigr)P_1\parg + \mu\ta ^2 r^2 g' P_2\parg\Biggr] \\
&- \partial_r\Biggl[\frac{\ta ^2br^2}{ \sqrt{\tX}}
\Bigl(\mu\frac{\dot{g}}{b}+\frac{f'}{\ta }\Bigr)P_1\parg
+  \mu\ta ^2r^2 \dot{g}P_2\parg\Biggr]=0,
\nonumber
\end{align}
and
\begin{align}
-& \partial_t\Biggl[\frac{\ta ^3 r^2}{\sqrt{{\tX}}}\Bigl(\frac{\dot{g}}{b}+\mu\frac{f'}{\ta }\Bigr)
P_1\parg
+\mu \ta ^2r^2 f' P_2\parg\Biggr]\nonumber\\
& +\partial_r\Biggl[\frac{\ta ^2br^2}{\sqrt{\tX}}\Bigl(\mu\frac{\dot{f}}{b}+\frac{g'}{\ta }\Bigr)P_1\parg + \mu\ta ^2 r^2\dot{f}P_2\parg \Biggr]\notag\\
&=\ta^2 b r \left[ P_0' +  \sqrt{\tX} P_1'+ W P_2'\right].
\label{eqn:eomg}
\end{align}
Eq.~(\ref{eqn:eomf}) has an exact solution given by 
$P_1(x_0)=0$.  Here
\begin{equation}
x_0 = \frac{ 1 + 6\alpha_3 + 12\alpha_4 \pm \sqrt{ 1+ 3\alpha_3 + 9\alpha_3^2 - 12 \alpha_4}}{3 (\alpha_3+4\alpha_4)} \label{gsol}
\end{equation}
unless $\alpha_3=\alpha_4=0$ in which case $x_0=3/2$.  Given $g= x_0 a r$,  Eq.~(\ref{eqn:eomg}) reduces to
\begin{equation}
\sqrt{X} = \frac{W}{x_0}+x_0,
\label{eqn:geqnsoln}
\end{equation}
and can be used to define solutions for $f$ (see \S \ref{sec:examples}).

For any such solution, the stress energy tensor
takes the cosmological constant form
\begin{equation}
T^{\mu}_{\;\,\nu} = -\rho \delta^{\mu}_{\;\,\nu} ,
  \end{equation}
  where
 \begin{equation}
 \rho = -p =\frac{1}{2} P_{0}(x_0).
 \end{equation}
 To restore physical units to the dimensionless stress tensor multiply by $m^2 M_{\rm pl}^2$.
 The Einstein equations (\ref{eqn:modeinstein}) then determine the metric functions $a$ and $b$ jointly in the presence
 of matter in a manner independent of the solution for  $f$.
The self-accelerating stress energy depends only on the universal form for the spatial \stucky\ field, $
g= x_0 a r$, common to the whole class of isotropic solutions.

\section{Field Fluctuations}
\label{sec:fluctuations}

In this section we derive a  complete set of equations of motion for the observable
properties of \stucky\ field fluctuations or equivalently
the field equation for  spatial \stucky\  fluctuations.   We then relate
 these results to the second order action and its dependence on the  temporal \stucky\ background solution and massive gravity parameters.

\subsection{Equations of Motion}

Now let us consider spherically symmetric \stucky\ perturbations.
Hereafter $f$ and $g$  represent the background solution 
for the \stucky\ fields, $a$ and $b$ the background metric, while
 $\delta f$ and
\begin{equation}
\Gs = \delta g - x_0  r \delta a,
\end{equation}
$\delta a$ and $\delta b$ quantify the perturbations.
Because our solutions are exact for any $a(r,t)$, the metric can have
perturbations away from a given background, e.g.\ the Friedmann-Robertson-Walker background,   due to perturbations in the matter sector and still be on the exact self-accelerating solution in the \stucky\ sector. 
 $\Gs$  therefore is the fluctuation in $g$ away from
the self-accelerating solution.

Linearizing Eq.~(\ref{eqn:eomf}) in the fluctuations, we obtain
a closed-form equation for $\Gs$
\begin{eqnarray}
0 &=& \dfeom \nonumber\\
&=&
\partial_t \left[ 
\frac{ a^2 r }{\sqrt{X}} \left( \frac{\dot f}{b} + \mu \frac{g'}{a} \right)\Gs \right]
 -\partial_r \left[
\frac{ab r }{\sqrt{X}} \left( \mu \frac{\dot g}{b} + \frac{f'}{a}\right)\Gs\right] 
\nonumber\\
&&\quad - \mu a^2 r^2  \left[ \frac{(a r)'}{ar } \dot \Gs -  \frac{\dot a}{a}\Gs' \right] .
\label{eqn:Gs}
\end{eqnarray}
 This equation for $\Gs$ contains no reference to matter or
metric perturbations.  It is also decoupled from the \stucky\ fluctuation $\delta f$.  The coefficients in this equation do however depend explicitly on the
background solution for the metric and both \stucky\ fields.   

This equation is first order in time and requires one 
set of initial data $\Gs(r,t=0)$ to solve.
However, note  that if we choose initial values of $\Gs(r,0)=0$, $\Gs$ will stay zero
regardless of matter-driven perturbations in the metric.   In this sense, 
the \stucky\ fluctuations are decoupled from the matter.

As was the case for the background, the perturbed Eq.~(\ref{eqn:eomg})
defines the evolution of $\delta f$ and depends explicitly on $\Gs$ as well as
on the metric fluctuations (see Appendix \ref{actionappendix}).   Thus to characterize the complete 
\stucky\ sector requires a second set of initial data $\delta f(r,0)$.  {Considered together, the amount of initial data and the two coupled first order systems are indicative of a single propagating degree of freedom.  
However, these fields are not interdependent in the usual way of a field and its conjugate momentum and in particular do not combine to form a single wave equation: $\Gs$ does not depend on
$\delta f$.} We will return to the
interpretation of these facts in \S \ref{sec:action}.

Even though the \stucky\ fluctuations are decoupled from the matter, a
finite $\Gs$ generates a metric perturbation, and hence a matter perturbation, due to
the effective stress energy it carries.
By explicitly evaluating
Eq.~(\ref{eqn:stressenergy}), we can write down 
the energy density perturbation
\begin{eqnarray}
\label{drhopert}
\delta\rho &=&  -\delta T^0_{\hphantom{0}0} 
=
 \left( \frac{  g'^2-f'^2+ a^2  W }{a^2  \sqrt{X}} - x_0 \right) \frac{P_1'\Gs}{2 a r } ,
 \label{eqn:deltarho}
 \end{eqnarray}
and the radial momentum density perturbation
\begin{equation}
\mom = T^r_{\hphantom{0}0} = - 
  \frac{ \dot{f}{f'}- \dot{{g}}{g'}}{a^2 \sqrt{X}}   \frac{P_1'\Gs}{2 a r} .
\end{equation}
Here and below the polynomials $P_i=P_i(x_0)$ are assumed to be evaluated
on the background.  
Note that both quantities are directly proportional to 
$P_1'$. Furthermore $\delta \rho$ and $\mom$ depend only on $\Gs$ and not its derivatives, in contrast to the energy and momentum of usual fields, which
typically also get contributions from the their kinetic energy.
%

The energy and momentum density  suffice to define the impact of the \stucky\ fields
on matter through the two associated Einstein equations.
To define the whole stress tensor or use all of the Einstein equations, we can also
derive the radial pressure
\begin{eqnarray}
\pr &=& - 
 \left( \frac{  \dot f^2 - \dot g^2
  +  b^2 W  }{ b^2 \sqrt{X}} - x_0\right)
  \frac{P_1'\Gs}{2 a r }. 
  \label{eqn:deltaprr}
  \end{eqnarray}
The other components of the stress energy tensor can then be found through
conservation of energy
\begin{equation}
\dot {\delta \rho} = - 3\frac{\dot a}{a} (\delta \rho + \delta p) + 
\frac{ (b a^3)'}{ba^3} \mom + \frac{ (r^2 \mom)'}{r^2}=0.
\end{equation}
Given $\delta \rho$ and $\mom$, this equation defines $\delta p$. 
Since by definition
\begin{equation}
\delta p = \frac{1}{3} (\pr + \pt + \pp)
\end{equation}
and $\pt = \pp$, the energy equation defines the remaining components. 
We have verified that direct evaluation of the stress components 
in Eq.~(\ref{eqn:stressenergy}) gives the same result once combined with the
equations of motion.
  With the stress tensor fully defined,
any two of the Einstein equations completes the dynamics as usual and define $\delta a$
and $\delta b$ jointly with the matter stress energy.  

Momentum conservation
provides another useful auxiliary equation
\begin{eqnarray}
\dot\mom &=& -\Bigl(5\frac{\dot{a}}{a} -\frac{\dot b}{b}\Bigr)\mom+\frac{b^2}{a^2}\Bigg[\frac{b'}{b}\Bigl(\delta \rho+\pr \Bigr)
\\&&
+  (\pr)'  + \frac{(ar)'}{ar}\left(  2\pr -\pt  -\pp \right) 
\Bigg].\nonumber
\label{eqn:mom}
\end{eqnarray}
With an anisotropic stress tensor, balancing radial gradients with the different $T^{i}_{\hphantom{i}j}$ components can result in static conditions
if
\begin{eqnarray}
 (\pr)'  =-
\frac{(ar)'}{ar}\left(  2\pr -\pt  -\pp \right) .
\label{eqn:stressbalance}
\end{eqnarray}
Whereas for an isotropic stress tensor, the right hand side vanishes and
$(\pr)' = \delta p'$, whose finite value would generate a momentum density unless balanced
by the $b'/b$ gravitational term. The impact of anisotropic equations of state on  hydrostatic equilibrium 
and stability 
has been studied in a related neutron star context (e.g.~\cite{1974ApJ...188..657B,1976A&A....53..283H}).

{
In summary,
similar to what we found for the background solution, 
the stress energy tensor for the fluctuations depends on the perturbation to the spatial \stucky\ field ($\Gs$)  and not
 the perturbation to the temporal \stucky\ field ($\delta f$).  Hence even though the \stucky\ fields require two sets of initial
data obeying  coupled first order equations, reminiscent of a field and its conjugate momentum, only one of these has any observable impact on the matter.  Moreover, the coupling is unidirectional, as $\Gs$ forms its own autonomous first order system.   It can be consistently set to zero by 
a choice of initial condition.
} 

\subsection{Action}
\label{sec:action}

In order to make contact with the literature on \stucky\ field dynamics, it is useful to see how these results arise from the second order action.  In particular, {apparent  differences} concerning the sign and value of the 
time derivative terms in this action have led to seemingly conflicting results about the presence of 
strongly coupled or ghost-like modes \cite{deRham:2010tw,Koyama:2011wx,Gumrukcuoglu:2011zh}.

We expand the action of Eq.~(\ref{eqn:genpot}) to second order in
both \stucky\ and metric perturbations around 
our background solution  in the Appendix.
It is sufficient here to examine the \stucky\ sector:
\begin{align}
S_{SS} = &\frac{P_1'}{2} M_{\rm pl}^2 m^2 \int dr d\Omega \Big[ \delta f \dfeom
- \frac{a b}{2}  \Gs^2 \nonumber \\
& -  a^2  b r \Gs ( \Gs'  \partial_{g'} +\dot\Gs \partial_{\dot g}) \sqrt{X}\Big] ,
\label{eqn:stuckyaction}
\end{align}
where $\dfeom$ was defined in Eq.~(\ref{eqn:Gs}) and note that this is the only term that
depends on $\delta f$ in the whole second order action (see Eq.~\ref{eqn:LagST}).
Furthermore since $\dfeom$ depends only on $\Gs$, 
there are no terms quadratic in $\delta f$ fluctuations.
Varying the action with respect to $\delta f$ reproduces the perturbed equation of motion $\dfeom=0$
{with no dependence on other perturbations}. 

With the help of Eq.~(\ref{eqn:geqnsoln}), we can see that the coefficient of the  $\delta f \dot\Gs$
term in the action is proportional  to
\begin{eqnarray} 
P_1'\left[  \frac{x_0 \dot f} {bW} - {\mu} \frac{(ar)'}{a} \right],
\label{eqn:kinetic}
\end{eqnarray}
and determines the dynamics of $\Gs$ through its equation of motion.
The quantity in brackets is determined by the background solution for $f$ and there is
a special case where it vanishes identically.  
Note that if we assume that the time dependence of $f \propto a^p$,
  then $\dot f \propto f \dot a/a$ and $b W \propto f \dot a/a$ so this condition becomes independent of
time. We shall see that this very special $f$-$g$ symmetry
is exploited in open universe solutions (see \S \ref{sec:fa}) and explains their lack of
\stucky\ dynamics   \cite{Gumrukcuoglu:2011zh}.

While $P_1'$ in Eq.~(\ref{eqn:kinetic}) drops out of the equation of motion, it does determine the
sign of  all \stucky\ coefficients in the action.  For the  self-accelerating solutions 
\begin{equation}
\label{p1p}
P_1'(x_0) = \pm4 \sqrt{1+3 \alpha_3+9\alpha_3^2 - 12 \alpha_4},
\end{equation}
for the two  $x_0$ solutions.   If $\alpha_3=\alpha_4=0$ there is only one solution, $P_1' = -4$ and the sign of  the coefficient is fixed entirely by
the background solution.
There is a special choice
\begin{equation}
\alpha_4 = \frac{1}{12} (1+ 3 \alpha_3 + 9 \alpha_3^2),
\end{equation}
where $P_1'=0$.  In this case the whole \stucky\ quadratic action vanishes as
do all of the stress-energy components.  This is the same special choice that was made in
Refs.\ \cite{Nieuwenhuizen:2011sq, Berezhiani:2011mt}, who similarly found that the quadratic
action for fluctuations around solutions with this parameter choice vanished.

 For general $\alpha_3$ and $\alpha_4$, $P_1'$  and hence the action may have either sign.
This is consistent with the decoupling limit analysis of the kinetic term where
$\alpha_3\ne0$ or $\alpha_4\ne 0$ was necessary to ensure that the helicity 0 scalar fluctuation is not a ghost \cite{deRham:2010tw,Koyama:2011wx}.   
In this limit, the scalar, $\delta\pi$, determines the two \stucky\ field fluctuations $\Gs \to \delta \pi'$, $\delta f \to -\dot {\delta \pi}$ and hence the $\delta f \dot \Gs$ term acts as a kinetic term for $\delta \pi$
after integration by parts (see also Eq.~\ref{eqn:pi}).

{
The decoupling limit form suggests a partial way to interpret the first order nature of the
coupled $\delta f$-$\Gs$ system.
There $\delta f$ appear in the action in a way similar to a canonical momentum for the field $\delta\pi$, which is itself related to $\Gs$
\footnote{We thank Andrew Tolley for this insight}. This viewpoint suggests an interpretation of $\delta f$ and
$\Gs$ as a pair of ``half" degrees of freedom.  This counting is related to the absence
of the Boulware-Deser ghost: of our original 4 \stucky\ fields, there is one constraint by construction and there are 
two angular modes that are absent in our analysis because of 
spherical symmetry.}

{However, interpreting $\delta f$ and $\Gs$ as a simple single propagating scalar 
does not carry over to the full theory, where the fields are not directly related.  This is of course not in itself a contradiction, as the structure of this theory makes a global helicity decomposition
generically impossible \cite{deRham:2011qq}.
In particular $\delta f$ has no necessary
relationship to $\Gs$ as any choice that solves its equation of motion yields
the same solution for $\Gs$.  Furthermore, the dynamics of $\Gs$
depends on the choice of background solution for $f$, as we shall see in the next
section.}

\section{Time Dynamics}
\label{sec:examples}

The detailed dynamics of \stucky\ field fluctuations depend on the choice of solution for
$f$, the \stucky\ field
related to the time coordinate.  
There are even special cases where the 
fluctuations have no dynamics.   
All such choices have the same background
metric and self-acceleration.
 In this section we  systematically construct explicit solutions for $f$ and study their implications for the dynamics of field fluctuations and the associated initial value problem.  We also generalize many solutions 
 in the literature \cite{Koyama:2011xz,Koyama:2011yg,Nieuwenhuizen:2011sq,deRham:2010tw,Koyama:2011wx,Gumrukcuoglu:2011ew} as well
 as introduce new classes.

\subsection{Unitary Gauges}

Since the \stucky\ fields are spacetime scalars, their values, once mapped to the same
spacetime point, are coordinate independent.   Unitary gauges are defined so their
time  $t_u$ and radial  $r_u$ coordinates are
\begin{eqnarray}
t_u = f(r,t) , \qquad r_u= g(r,t).
\label{eqn:mapping}
\end{eqnarray}
The \stucky\ fields are therefore just the unitary gauge coordinates  and their functional form in 
isotropic coordinates $(r,t)$ is simply the map itself.   In unitary gauges, 
the fiducial metric $\Sigma_{\mu\nu}=\eta_{\mu\nu}$ and this simplification has
been crucial for finding most solutions existing in the literature.  While we do not utilize a unitary gauge construction in our solution,
 we will categorize solutions by their unitary gauge  correspondence.

All self-accelerating solutions have
the common property that $g= x_0 a r$, implying that the radial coordinate of any unitary
gauge  is simply the same conformally rescaled version of the isotropic radial coordinate, similar
to the distinction between physical and comoving coordinates in cosmology.

The different solutions are thus characterized by the choice of time coordinate for the unitary gauge.
Each of these choices solves Eq.~(\ref{eqn:geqnsoln})  for the same $g$; written explicitly,
\begin{eqnarray}
&&  b^2  f'^2  + 2 a r(  a' \dot f^2 - \dot a \dot f f' )  + r^2(a' \dot f - \dot a f')^2  
  = 
\nonumber\\
&&\qquad  x_0^2 ( a'^2 b^2  r^2 + 2   a'  a b^2 r - \dot a^2 a^2 r^2  ).
\label{eqn:fsoln}
\end{eqnarray}
Recall that the metric functions $a$ and $b$ do not depend on the solution for
$f$ and so may be considered external functions in solving this equation.

There are several properties of this equations that are useful to note. 
It is a nonlinear partial differential equation in space and time.  Its
solutions can be characterized by the boundary conditions
$f(0,t)$ and $f(r,0)$.   The former is equivalent to the local $r=0$ relationship between $t$ and unitary time $f$ and is our primary classification criterion.
  The latter allows a family of such solutions associated
with a time integration constant.
{The equation does not guarantee that the map in Eq.~(\ref{eqn:mapping}) is free from
singularities everywhere in the spacetime.   A singularity corresponds to a change in the sign of the
Jacobian determinant $\dot f g' - \dot g f'$ and hence a change in the sign of $\mu$ 
(see Eq.~\ref{eqn:XW}).   Solutions should be checked against singularities 
developed in the course of radial or temporal integration as they signal a breakdown in
our treatment of the square roots in the action.}

 Finally given a solution 
for the $\alpha_3=\alpha_4=0$ case where $x_0=3/2$,
\begin{equation}
f(r,t;\alpha_3,\alpha_4) = \frac{2 x_0}{3} f(r,t;0,0) 
\end{equation}
is the solution for the full parameter space.  One may similarly scale
any other specific  $\alpha_3$ and $\alpha_4$ case to the general case by use of
the ratios of the respective values of $x_0$.  This relationship makes it trivial
to generalize special solutions in the literature (cf.~\cite{Koyama:2011yg}).

In our examples, we seek cosmological solutions where
\begin{eqnarray}
b(r,t) = 1, \qquad 
a(r,t) =  \frac{\fa(t)}{1+K r^2/4},
\end{eqnarray}
where $K$ is the spatial curvature and $\fa(t)$ is the scale factor of
a Friedmann-Robertson-Walker metric.  Note that in a flat $K=0$ geometry
$a(r,t)= a(t)= \fa(t)$.

\subsection{Unitary Time $t$}
\label{sec:ft}

Perhaps the most natural choice of solution is to take unitary time $f$ to
be linearly related to $t$ 
\begin{equation}
f(0,t) =\frac{x_0}{C}  t ,
\end{equation}
where $C$ is a constant.
For a flat universe $K=0$ and
 we can solve Eq.~(\ref{eqn:fsoln}) with this ansatz
to ${\cal O}(r^4)$ 
\begin{equation}
f(r,t) = \frac{x_0}{C H} \left( H t + \frac{1 - \sqrt{1-C^2}} {2} ( a H r)^2 \right),
\end{equation}
where $H =\dot a/a$.  Note that on cosmological scales where $a H r \gtrsim 1$, this temporal field becomes spatially inhomogeneous.  

\subsubsection{de Sitter}

The special case that $H={\rm const.}$  is of course the late
time limit of the self-accelerating solution where the matter becomes subdominant
to the effective cosmological constant
\begin{equation}
H = \sqrt{ \frac{1}{6} P_0} m ,
\end{equation}
and we can normalize $a = e^{H t}$.
Solving Eq.~(\ref{eqn:fsoln}) for this case with the ansatz
\begin{equation}
f(r,t) = \frac{x_0}{C H} \left[ H t + F( a H r) \right]
\end{equation}
gives
\begin{eqnarray}
f(r,t) &=& \frac{x_0}{ CH }\left[ H t -  y+  \ln \Big| \frac{1+y} { 1 - (a H r)^2}\Big|
\right],
\nonumber\\
y&=& \sqrt{1+ C^2( a^2 H^2 r^2 -1)} .
\label{eqn:koyamasoln}
\end{eqnarray}
These relations are equivalent to the de Sitter solutions of Refs.~\cite{Koyama:2011xz,Koyama:2011yg} up to an overall constant and generalizes them to 
arbitrary $\alpha_3$, $\alpha_4$.

Note that if $C=1$, $f\propto t$ and $g \propto ar \equiv r_p$ near $r=0$
and can
be thought of deriving from a Lorentz scalar in physical coordinates
\begin{equation}
\pi = \frac{1}{2} x_0(r_p^2- t^2) 
\label{eqn:pi}
\end{equation}
with $ f \approx -  \dot \pi$ and $g \approx  \partial_{r_p} \pi$. Hence $C\ne 1$ {may be} said to have
a ``vector" in the background   \cite{Koyama:2011yg} even though the \stucky\ fields are spacetime
scalars.

Given the solution in Eq.~(\ref{eqn:koyamasoln}), the 
 evolution equation for $\Gs$ becomes
 \begin{eqnarray}
(y^2+y)\frac{ \dot \Gs}{H} &=& \left[y+ \frac{y^2}{(a H r)^2}\right] r \Gs'  \nonumber\\
&& + \left[\frac{2 y^2}{(a H r)^2} - 3 y^2 + 1 \right] \Gs,
\end{eqnarray} 
 and the energy density, momentum density and pressure are given by
\begin{eqnarray}
\delta\rho &=& -
\frac{ (a H r)^2  C^3}{(1+C) (1-y)^2}
\frac{x_0 P_1' \Gs }{2 a r}, \nonumber\\ 
\mom &=& \frac{y+ C^2-1 }{C^2 a^2 H r} \delta\rho,\quad
\delta p = \frac{1}{3}\delta \rho \,.
\end{eqnarray}
 The pressure follows a radiation-like equation of state but
itself is composed of anisotropic contributions
\begin{equation}
\pr  =  \left( \frac{y+ C^2 -1 }{ C^2 a H r} \right)^2
\delta\rho \ne \delta p.
\end{equation}
 It is instructive to consider
the simplest $C=1$ case further.  Here $y=a H r$, and given an initial
value for $\Gs(r,0)$, the general solution is
\begin{equation}
\Gs(r,t) =\frac{1}{a^3} \left(1 +  \frac{a-1}{a H r } \right)^3 
\Gs\left(r + \frac{a-1}{a H},0\right),
\end{equation}
which then defines  the stress components
\begin{eqnarray}
\delta\rho &=& - \frac{( a H r)^2}{2 ( 1-  a H r)^2} \frac{x_0 P_1'\Gs}{2 a r} ,\nonumber\\
\mom &=&  \frac{1}{a} \delta \rho ,\quad \pr = 3 \delta p = \delta \rho .
\end{eqnarray}
The angular stresses vanish and the total pressure is given 
by the radial component.  {These components have poles in their
expressions at the horizon $r=1/aH$ corresponding to the coordinate singularity in
Eq.~(\ref{eqn:koyamasoln}) for unitary time.}    However, since the solution for $\Gs$
always maps the initial horizon onto the horizon at a later epoch,  if the
energy density is finite in the initial conditions there, it remains so.
Furthermore for large scales $r\gg 1/H$, {which were} outside the horizon at the initial epoch,  $\delta \rho \propto a^{-4}$; 
 this is  consistent with expansion effects and the radiation-like equation of state.

\subsubsection{Matter and Radiation Domination}
\label{sec:koyamaw}

 While we have solved the initial value
problem for fluctuations in the de Sitter limit for $f(0,t) \propto t$, in a  cosmological solution these
would themselves originate from the dynamics in the preceding radiation and matter dominated
epochs.

To construct solutions where $f(0,t) \propto t$ during the matter and radiation epochs, we first try
\begin{equation}
f(r,t) = \frac{x_0 t}{C}  F(a r/t) ,
\end{equation}
with the assumption that $H \propto a^{-3(1+w)/2}$ for a constant $w$ and $C$.
For any such choice, 
we can solve differential equation for $F$ implied by Eq.~(\ref{eqn:fsoln}).
While this construction would seem to admit many solutions, 
for finite $C$,  $\mu$ changes sign 
between $+1$ for $r\rightarrow 0$ and $-1$ for $r\rightarrow \infty$ indicating 
a singularity in the map between
isotropic and unitary coordinates.  

On the other hand, the $C \rightarrow \infty$ limit does not suffer this problem
and yields
\begin{equation}
f(r,t) = {x_0  }\sqrt{ \tau^2(t) + {r^2 a(t)^2} },
\label{eqn:fconstw}
\end{equation}
where
\begin{equation}
\Tt =\sqrt{\frac{9(1+w)^2}{4(2+3w)} }\,\, t
\label{eqn:Ttt}
\end{equation}
for $w>-2/3$.  Eq.~(\ref{eqn:Ttt}) is now easy to generalize to 
an arbitrary expansion history by matching
\begin{equation}
\Tt^2(t) = a(t)\int^t \frac{d\tilde t}{\dot a(\tilde t)} 
\end{equation}
for $\dot a >0$.
This solution includes a universe that evolves from radiation domination through matter
domination to self-acceleration.  Note that since $w=-1$ during self-acceleration,
$\Tt$ is no longer directly related to $t$ and so this solution is distinct from the solutions of
the previous section.  

The fluctuation $\Gs$ evolves under 
\begin{eqnarray}
 2 r^2 a^3 \dot a \dot \Gs + [\dot a^2  \Tt^2 - a^2] r \Gs' =-2\frac{A}{B} \Gs ,
\end{eqnarray}
where
\begin{eqnarray}
A &=&
\dot a^4 \Tt_r^4  + 2 a \dot a^3   \Tt_r^3 + r^2  a^2 \dot a^4
\Tt_r^2  - 2 a^3 \dot a \Tt_r (1 - r^2 \dot a^2) \nonumber\\
&&+ 2 r^2 a^5 \ddot a -
a^4 (1 - \dot a^2  r^2 + 2 \dot a^4 r^4 - 2 \dot a \ddot a r^2 \Tt_r) ,\nonumber\\
B &=& ( a + \dot a \Tt_r)^2 - r^2 a^2 \dot a^2,
 \end{eqnarray}
with $\Tt_r^2 = \Tt^2 + r^2 a^2$ and the stress energy components are
\begin{eqnarray}
\delta \rho &=& - \frac{ a^2 \dot a r}{B}
\frac{x_0 P_1' \Gs}{ \Tt_r} , \,\,\,
\mom = \frac{a^2 - \dot a^2 \Tt^2 }{2 r a^3 \dot a} \delta \rho, \nonumber\\
\delta p &=&  \frac{ a \ddot a}{3 \dot a^2} \delta \rho, \,\,\,
\pr = \left( \frac{a^2 - \dot a^2 \Tt^2 }{2 r a^2 \dot a}  \right)^2\delta \rho .
\end{eqnarray}
Note that for a constant {background} equation of state {$w$}, $\delta p/\delta \rho = -(1+3w)/6$.
For $w>-1/9$ expansion thus makes the energy density redshift more slowly than
the dominant matter but always faster than the constant background \stucky\ contributions.  Thus in linear theory starting from some arbitrary field configuration,
  we expect that
$\delta \rho/\rho$ will be driven rapidly to zero for any matter content leaving just
the cosmological constant background.

\subsection{Unitary Time $a$}
\label{sec:fa}

Unitary time can also be made proportional to $a(0,t)=\fa(t)$ 
\begin{equation}
f(0,t) =\frac{x_0}{C} \fa(t),
\end{equation}
where, again, $C$ is a constant.
In this case both $f$ and $g$ share the same temporal dependence at
the origin and hence provide the ingredients necessary for static solutions.   
To order ${\cal O}(r^4)$ 
\begin{equation}
f(r,t) \approx \frac{x_0}{C}\fa \left[ 1 + 
\frac{ {\fadot}^2 \pm \sqrt{ (\fadot^2 - C^2)(\fadot^2+K)}}{2}r^2 \right].
\label{eqn:feqna}
\end{equation}
Here we have kept the possibility that $K\ne 0$ since it allows a 
special class of solutions.

\subsubsection{Open Solution}

If $K<0$ then Eq.~(\ref{eqn:feqna}) has a simple solution for $C^2=-K$.  
Taking this ansatz, the full solution is \cite{Gumrukcuoglu:2011ew,Gumrukcuoglu:2011zh}
\begin{equation}
f(r,t) =x_0 \fa(t) \sqrt{ \frac{1}{-K} + \frac{r^2}{(1+K r^2/4)^2}}.
\end{equation}
Note that the open solution for $f$, like other solutions, is inhomogeneous in isotropic
coordinates.  There is nothing special about the open solution with regards
to homogeneous and isotropic \stucky\ fields (cf.~\cite{Gumrukcuoglu:2011ew}).

On the other hand, the common separable $\fa(t)$ factor in $f$, $g$ and $a$ allows this solution to satisfy
the static condition for the $\Gs$ field, Eq.~(\ref{eqn:kinetic}).  Correspondingly, Eq.~(\ref{eqn:Gs})
becomes
\begin{equation}
(4 + K r^2) r\Gs' + 2(4-Kr^2) \Gs =0 .
\end{equation}
Note that $\mu ={\rm sgn}(\fadot)$ and so is $+1$ for an expanding universe.
Interestingly the determinant goes to zero if the expansion turns around, e.g.~because of a negative true cosmological constant, signaling a breakdown of the solution.   We therefore consider only expanding solutions here.

The general solution to this equation is
\begin{equation}
\Gs(r)\propto\left(\frac{ 4 + K r^2}{r} \right)^2 .
\end{equation}
This static \stucky\ field produces no energy density, momentum
density or pressure; i.e.,
\begin{equation}
\delta p = \mom = \delta p = 0 .
\end{equation}
These static conditions are maintained by a delicate balance of the radial
and anisotropic stress gradients
\begin{equation}
\pr =\left( 1 - \frac{ \fadot}{\sqrt{-K}} \right) \frac{x_0 P_1' \Gs}{2 ar} ,
\end{equation}
which satisfy the static condition for Eq.~(\ref{eqn:stressbalance}).

{Even though $\delta p=\mom=0$, anisotropic stress still has an impact on the metric
through the Einstein equations.   Moreover,
these types of solutions are potentially unstable to anisotropic perturbations
\cite{DeFelice:2012mx,Gumrukcuoglu:2012aa}.}

\subsubsection{Flat Solution}

If $K=0$, then there is still one simple solution to Eq.~(\ref{eqn:feqna})
 where
$C^2 \ll \fadot^2$.   In that case
\begin{equation}
f(r,t) \approx \frac{x_0}{C} a(t) .
\end{equation}
To promote this to an exact solution we follow Ref.~\cite{D'Amico:2011jj} and generalize their approach
to arbitrary $\alpha_3,\alpha_4$
\begin{equation}
f(r,t) = \frac{x_0}{C} a(t) \left[ 1+ \frac{C^2}{4} r^2 + \frac{C^2}{4 } \frac{\Tt^2(t)}{a^2(t)}\right],
\end{equation}
which recovers the approximate scaling for $r  \ll 1/C$ and $t\ll a/C$.
Eq.~(\ref{eqn:Gs})
becomes
\begin{equation}
- 2 C^2 r^2 a \dot a \dot \Gs
+ [C^2 + (C^2 r^2-4) \dot a^2]r \Gs' =2\frac{A}{B}\Gs
\end{equation}
with
\begin{eqnarray}
A&=&
\mu C^4 +4 C^3 \dot a -16 C \dot a^3 -\mu (16- C^4 r^4) \dot a^4 \nonumber\\
 && -
2 C^3 r^2 a (C\mu + 2 \dot a) \ddot a,  \nonumber\\
B &=& {-\mu C^2 -4 C \dot a - \mu(4 - C^2 r^2) \dot a^2} .
\end{eqnarray}
Despite solving Eq.~(\ref{eqn:fsoln}), there is a problem with this form as a global
solution since
\begin{equation}
\mu = 
\begin{cases}
1 & a H r< \sqrt{1+ 4H^2/C^2} \\
-1 & aH  r>   \sqrt{1+ 4 H^2/C^{2}}
\end{cases}.
\end{equation}
The sign change in $\mu$
indicates the mapping to unitary gauge is singular unless $C \rightarrow 0$.

In this limit we are again left with a static equation
\begin{equation}
-  r \Gs' = 2 \Gs ,
 \end{equation}
 whose solution is $\Gs \propto 1/r^2$.   Thus the $K\rightarrow 0$ flat limit of the open solution is the same as the $C\rightarrow 0$ solution of the flat solution.  More generally, the stress-energy components
 become
\begin{eqnarray}
\delta\rho &=&   \frac{ C^3 r^2 \dot a} {B}
\frac{x_0 P_1' \Gs}{2 a r}, \,\,\, 
\mom = \frac{C^2 - (4-C^2 r^2) \dot a^2}{ 2 C^2 r a \dot a} \delta \rho, \nonumber\\
\delta p &=& \frac{ a \ddot a}{3 \dot a^2} \delta\rho, \,\,\, 
\pr  = \left[ \frac{C^2 - (4-C^2 r^2) \dot a^2}{2 C^2 r \dot a} \right]^2\delta\rho  .
\end{eqnarray}
Again for the $C\rightarrow 0$ case, $\delta \rho$, $\mom$, $\delta p$ are all suppressed whereas $\pr$, $\pt$, $\pp$ are unsuppressed and contain radial gradients
that delicately balance each other.
 
\subsection{Unitary Time $\sqrt{a}$}

While $f(0,t) \propto t$ or $a$ are obvious choices for solutions, they do not
exhaust the possibilities.    We have seen that the former generally require dynamical
\stucky\ fluctuations whereas the latter admit static ones.    However those static
cases are only formally well-defined as global solutions for $K<0$.    

To see whether there exists a well-defined static solution for $K=0$ we construct
solutions where 
$f(0,t) \propto \sqrt{a}$ is the limiting form.   Similarly to the
flat $\propto a$ case, we find an exact flat solution
\begin{equation}
f(t,r) = x_0  \sqrt{ \frac{a(t)}{C^2}  + \Tt^2(t) +  a^2(t) r^2 }
\end{equation}
for any constant $C$.  Note that if $C^2\gg \Tt^2/a $, this solution 
recovers the general $f(0,t) \propto t$ solution of \S \ref{sec:koyamaw}.
Since we have already considered the general case of this limit, it is
interesting to consider the opposite one.
Note that in a cosmological solution, once self acceleration sets in $\Tt^2/a
\rightarrow$ const.\ of order $H^{-2}$.  Thus taking $C \ll m$ ensures that
this limit is satisfied for all time.  

In this $C \ll m$ and $a=e^{Ht}$ case, to leading order $\Gs$ obeys 
\begin{equation}
{2 a r^2 C^2}\frac{\dot \Gs}{H}+ r \Gs' = -2 \Gs,
\end{equation}
which is a stiff equation as $r \rightarrow 0$ with an equilibrium static
solution of $\Gs \propto 1/r^2$.
The stress components
\begin{eqnarray}
\delta \rho &=& -\dfrac{r^2 C^3}{H}
\dfrac{ x_0 P_1'\Gs}{ \sqrt{a} r} , \,\,\, 
\mom = -\frac{ H }{2 a r C^2} \delta\rho , \nonumber\\
\delta p &=& \frac{1}{3}\delta \rho , \,\,\,
\pr =\frac{ H^2}{4 r^2 C^4} \delta\rho.
\end{eqnarray}
Since $C \ll H$, $\pr \gg \delta p$ and again involves a delicate balance of
anisotropic stresses.   
This case is very similar to
the $C\rightarrow 0$ case of the previous section but has the benefit
that $\mu=1$ everywhere and so $C$ can be set to a finite number.

\section{Discussion}
\label{sec:discussion}

We have presented a general analysis of isotropic \stucky\ field perturbations
around the full class of self-accelerating solutions \cite{Gratia:2012wt} of the massive gravity theory.   These background solutions are defined by two fields, one spatial $g(r,t)$ and one temporal $f(r,t)$, 
where only the spatial one is responsible for the stress-energy of the effective cosmological
constant.

Likewise \stucky\ field perturbations come in two classes, spatial ($\Gs$) and temporal ($\delta f$).   {Spatial perturbations} can be consistently set to zero by a choice of initial conditions.  They are not generated   by any matter-induced metric perturbations as was already apparent
in that the self-accelerating solution is exact and non-perturbative for isotropic metrics  \cite{Gratia:2012wt}.
With an arbitrary choice of initial conditions, spatial \stucky\ fluctuations generically possess
stress energy. This stress energy produces metric fluctuations to which the matter
responds.     Temporal \stucky\ fluctuations
carry no stress energy and have {no} {effect on} metric fluctuations at this order
in perturbation theory.

Importantly, the dynamics of spatial \stucky\ fluctuations 
 make  energy density deviations from the constant background always decay with the expansion in cosmological solutions,
 implying that the background solution is stable.
There are special choices where the \stucky\ fields perturbations are
static,  but in those cases the fields carry no energy density.   In the general case,
the decay rate depends on the equation of state in the background.   

Different behaviors of the  spatial \stucky\ fluctuation are related to the {background} temporal \stucky\ field. 
This difference is the source of apparently conflicting claims in
the literature regarding \stucky\ perturbation dynamics.    The temporal \stucky\ field
in the background is itself the choice of the time coordinate in which the massive gravity
theory appears locally as the linearized Fierz-Pauli theory, i.e.\ it is a choice of unitary gauge 
for the covariant theory.  For unitary time coinciding locally with isotropic time, the
single propagating degree of freedom found in the decoupling limit  \cite{deRham:2010tw,Koyama:2011wx} appears in the exact theory as well.  For unitary time that scales
with the spatial scale factor, there are no \stucky\ dynamics, consistent with the open universe 
solutions  \cite{Gumrukcuoglu:2011zh}.  Furthermore these solutions remain static not by
possessing a  vanishing stress tensor perturbation but rather by a potentially unstable balance of 
anisotropic stresses.

The sign of the energy density carried by the \stucky\ field $\Gs$, as well as the sign of the prefactor of the term in the quadratic action that generates the $\Gs$ dynamics, are both determined by the quantity $P_1'$, a constant determined by the parameters of the theory ($\alpha_3$ and $\alpha_4$, see Eq. \ref{p1p}).
 $P_1'=-4$ when $\alpha_3=\alpha_4=0$, but can be either positive or negative for 
general $\alpha_3$ and $\alpha_4$. This is consistent with  decoupling limit findings  that perturbations around self-accelerating solutions generically are ghost-like for $\alpha_3=\alpha_4=0$,
but can be made healthy in more general circumstances \cite{deRham:2010tw,Koyama:2011wx}. $P_1'$ can be also be set to zero by a special
choice.   These models have no quadratic \stucky\ contributions to the action and no linearized stress energy perturbations, 
consistent with  the findings of Ref.~\cite{Nieuwenhuizen:2011sq, Berezhiani:2011mt}, who also made this parameter choice.

In the exact theory, the dynamical system represented by the \stucky\ fields exhibits some 
peculiar properties that are obscured in decoupling limit analyses.    
The two \stucky\ field perturbations require two sets of initial data -- $\delta f(r,0)$ and $\Gs(r,0)$ -- and each obey coupled first order differential equations of motion.
 This is the amount of initial data and dynamics that
we would expect for a single propagating degree of freedom.  This situation might have been anticipated at the outset from a counting argument.  Starting with 4 \stucky\ scalar degrees of freedom, we remove two by restriction to spherical 
symmetry, and a third (the Boulware-Deser ghost) is removed by construction in this theory. 

However, these two
``half" degrees of freedom are unusual in their interrelation. As indicated above, the $\Gs$ field has a first order
equation of motion that is independent of the $\delta f$ field. The $\delta f$ dynamics do depend on the $\Gs$ field,
but they carry no stress energy at this order in perturbation theory.
 In addition to this, we find that the time derivatives of $\Gs$ 
do not actually contribute energy or momentum density either, so that there is no obvious classical instability associated with
these fields even when they are apparently ghost-like.

Finally, we uncovered a potential physical problem with certain background configurations for the temporal
\stucky\ field.   The structure of the theory itself does not prevent the determinant of the fiducial metric from vanishing when solving the temporal or spatial boundary value problem. 
Some solutions for $f$ pass through a zero determinant at finite 
radius (e.g.\ \cite{D'Amico:2011jj}) or time from well-posed initial values.   At this point, the mapping between unitary and isotropic coordinates becomes singular.   

Since such a choice may have physical pathologies, we also constructed a new solution for $f$ that
passes from radiation domination through matter domination to self-acceleration without exhibiting this potential problem or resorting to static fields.   We leave the larger question of the theoretical implications for the existence of
potentially pathological solutions to a future work.

\smallskip{\em Acknowledgments.--}  We thank Peter Adshead, A. Emir Gumrukcuoglu, Daniel Holz, Claudia de Rham, Kazuya Koyama,  and Andrew Tolley for helpful discussions. 
MW and WH were supported by U.S.~Dept.\ of Energy contract DE-FG02-90ER-40560. WH was additionally supported by Kavli Institute for Cosmological Physics at the University of Chicago through grants NSF PHY-0114422 and NSF PHY-0551142  and an endowment from the Kavli Foundation and its founder Fred Kavli and by the David and Lucile Packard Foundation.  PG was supported by the National Research Fund Luxembourg through grant BFR08-024. 

\hfil
\onecolumngrid
\appendix

\section{Second Order Action}\label{actionappendix}

The quadratic Lagrangian for the perturbations ($\Gs$, $\delta f$, $\delta a$, $\delta b$) 
can be separated out as
\begin{equation}
\delta^2 {\cal U}_g = \delta^2_{SS}{\cal U}_g+  \delta^2_{ST} {\cal U}_g +   \delta^2_{TT} {\cal U}_g,
\end{equation}
where ${\cal U}_g = \sqrt{-g}{\cal U}/(4 \sin\theta)= a^3 b r^2 {\cal U}/4 $ carries the relevant pieces of the
metric determinant,
 $S$ refers to pieces involving  \stucky\ fluctuations and $T$ those involving 
metric fluctuations.    The \stucky-\stucky\ terms are
\begin{eqnarray}
  \delta^2_{SS} {\cal U}_g &= &\half \Gs^2 a b \( P_0''+P_1'' \sqrt{X} +P_2'' W\)
+ a^2 b r P_2' \Gs \left( \Gs' \partial_{g'}W + \dot \Gs \partial_{\dot g} W \right) \nonumber\\
&&+ a^2 b r  P_1' \Gs \left[ \Gs' \partial_{g'}  +  \dot \Gs\partial_{\dot g}   + \delta \dot f \partial_{\dot f}+ \delta f' \partial_{f'}  \right] \sqrt{X}
+  \mu a r(2P_2 - x_0 P_2') \delta f  (\dot \Gs\partial_r - \Gs' \partial_t)(  a r )
\nonumber\\
&=&  P_1' \left[
- \delta f \dfeom +\frac{ab}{2}  \Gs^2 +   a^2  b r  \Gs
(   \Gs'  \partial_{g'}
+   \dot\Gs \partial_{\dot g}   ) \sqrt{X} \right],
\end{eqnarray}
where we have used the background solution $g=x_0 a r$, the polynomial recursion Eq.~(\ref{eqn:recursion}), and
the relationship between $W$ and $\sqrt{X}$ on the background from Eq.~(\ref{eqn:geqnsoln}).  Note that equality here means equality up to total derivative terms.  Here
\begin{eqnarray}
\dfeom \equiv
\partial_t [ a^2 r b \Gs \partial_{\dot f} \sqrt{X}]+
\partial_r [ a^2 r b \Gs \partial_{f'} \sqrt{X}]
 - \mu a r  [ (a r)'  \dot \Gs -  \dot a r \Gs' ] .
\end{eqnarray}
and its correspondence to Eq.~(\ref{eqn:Gs}) can be established by noting
\begin{equation}
\partial_{\dot f} \sqrt{X} = \frac{1}{b\sqrt{X}} \left( \frac{\dot f}{b} + \mu\frac{g'}{a} \right),
\qquad
\partial_{f'}\sqrt{X} = -\frac{1}{a\sqrt{X}} \left( \mu \frac{\dot g}{b} + \frac{f'}{a} \right).
\end{equation}

Next, the mixing between the \stucky\ fields and the metric fluctuations is given by
\begin{eqnarray}
\delta^2_{ST} {\cal U}_g  &=& abr \left[ 2 P_0' +P_1'  (2+a \partial_a+ a x_0 \partial_{g'})\sqrt{X}
+ P_2' (1 +a x_0 \partial_{g'})W \right] \delta a \Gs 
 \nonumber\\
&& +  
a^2 b r^2 x_0 \left[ (P_2' \partial_{\dot g} W +  P_1' \partial_{\dot g}\sqrt{X})  \dot {\delta a} +
 (P_2' \partial_{g'} W +  P_1' \partial_{ g'}\sqrt{X}){\delta a'} \right] \Gs \nonumber\\
&&+  2 a^2 b r^2  P_2  \delta a  (\dot \Gs \partial_{\dot g} +    \Gs'\partial_{g'} )W  + a^2 r
\left[ P_0' + P_1'(1+ b\partial_b) \sqrt{X} \right]  \delta b \Gs 
\label{eqn:LagST}
\\
 &=& a^2 br  P_1'  \Big\{ \left[  (1+a \partial_a)\sqrt{X} - x_0 \right] \frac{\delta a}{a} 
+  
  \left[{(  \delta a\,r)'}  \partial_{ g'} +
  \dot {\delta a}\, r \partial_{\dot g} \right](x_0 \sqrt{X}  -W)
+ 
\left[ (1+ b\partial_b) \sqrt{X} - x_0 \right]  \frac{\delta b}{b} \Big\}\Gs .
\nonumber
\end{eqnarray}
Both the $SS$ and $ST$ pieces are directly proportional to $P_1'$.
Note that 
\begin{eqnarray}
(1+ a\partial_a) \sqrt{X}  =  \frac{\dot f^2 - \dot g^2 + b^2 W}{b^2 \sqrt{X}}, \qquad
(1+b\partial_b)\sqrt{X} = \frac{  g'^2-f'^2+ a^2  W }{a^2  \sqrt{X}} ,
\end{eqnarray}
and these factors appear also in the perturbed energy and radial pressure relations,
Eqs.~(\ref{eqn:deltarho}) and (\ref{eqn:deltaprr}),
 since functional variation of the action with the metric must return these
components.
Note that there is no mixing between the $\delta f$  and metric fluctuations, as anticipated from our direct calculation of the stress energy tensor.  On the other
hand, variation of the action with respect to $\Gs$ yields an equation of motion for
$\delta f$ that depends on all of the other perturbations.

Finally, the metric-metric part is 
\begin{align} 
 \delta^2_{TT}{\cal U}_g= & 3 a r^2  P_0 \left( b \delta a^2  + a \delta a \delta b \right).
\end{align}
As expected, it represents the perturbation to $ \sqrt{-g}$ multiplying the effective background cosmological constant.
%
%
%

\bibliography{paperbib}

\end{document}